\documentclass[aps,prb,twocolumn,showpacs,groupedaddress]{revtex4}
\usepackage{graphicx,amsmath,verbatim}

\begin{document}

\title{Spin polarization effects in a two dimensional mesoscopic electronic structure with 
Rashba spin-orbit and lateral confinement}
\author{Stefano Chesi}
\author{Gabriele F. Giuliani}
\affiliation{Department of Physics, Purdue University,
West Lafayette, IN 47907, USA}

\date{\today}

\begin{abstract}
Because of the peculiar coupling of spatial and spin degrees of freedom 
effected by the Rashba spin-orbit interaction, lateral confinement of a two 
dimensional electronic system leads to a finite transverse spin polarization 
near the longitudinal edges of a current carrying quantum wire. The sign of this
component of the polarization is opposite at the two edges and can be reversed 
upon inversion of the current. Interestingly for small spin orbit coupling this
is the largest contribution to the total polarization, its magnitude being of 
second order in the coupling constant. As a consequence this phenomenon cannot 
be revealed in lowest order perturbative approaches. An in plane spin polarization
component is also present that is perpendicular to the current. Within the same
model this component would be also present in the bulk. On the other hand while 
in the latter case its magnitude is linear in the coupling constant, 
we find that it only represents a third order effect in the wire geometry. 
Our results are consistent with a general rigorous argument on the parity of 
the components of the spin polarization with respect to the sign of the spin 
orbit coupling constant.
\end{abstract}


\maketitle

We consider in this paper the effects of the lateral confinement on the properties of
a two-dimensional (2D) electron liquid in the presence of linear Rashba spin orbit.
The basic idea underlying our work stems from the simple observation that in order
to construct a localized wavefunction one must superimpose plane waves with opposite 
components of the wave vector along the direction of localization. 
Because spin-orbit forces the latter to have opposite spin components along the unrestricted
direction, it is clear that in the direction of confinement the electronic states must display 
a correspondingly inhomogeneous spin polarization whose direction also changes in space. 
In the particular case in which a 2D electronic paramagnetic system is laterally confined 
to produce a quantum wire, the superposition involves spins that are parallel to the 
plane of motion and therefore combine to give an out of plane polarization component.
Since the sign of this component of the magnetization is opposite for states 
with opposite momenta along the wire, it is clear that a net polarization can be established 
when the occupation of such states is unbalanced as in the case in which a net steady state 
current is forced through the system. As we will show this net perpendicular static
spin polarization is most pronounced near the lateral boundaries and displays an 
oscillatory behavior. Inverting the direction of the current simply flips the polarization.
No spin current is involved in the problem.

In order to demonstrate and exemplify this idea, we present here a complete analysis of
the simplest possible model hamiltonian describing a system of non interacting electrons 
subject to a homogeneous neutralizing background, a linear Rashba spin orbit coupling and 
a lateral confining potential.\cite{commQwire0}
The effect of impurities is neglected although impurity 
scattering in the channel will not change our main conclusions.
Although more realistic model potentials could be readily employed we have moreover chosen 
to describe the latter by means of infinite potential barriers, an assumption that allows 
us to present a transparent and explicit discussion while at the same time affording us a 
full description of the physical phenomenon.
A great advantage of our model is that an analytic perturbative expansion in the spin-orbit
coupling constant can be developed, a procedure that shows how the effect would not
be revealed by lowest order perturbation theories.
A generalization to more general smooth gating potentials is readily implemented.

That a steady state current can lead to a spin polarization in a 2D electron liquid subject
to Rashba spin orbit coupling is hardly unexpected. It is in fact well known that a current
forced along the $x$ direction of a bulk 2D electronic system will be accompanied by an in 
plane polarization along $y$.\cite{aronov89,aronov91,edelstein90,kato04a,silov04}
For small spin orbit this polarization is linear in the coupling constant. 

As we shall see however, while this component of the polarization is still present, 
confinement leads to a spin polarization that, for small spin orbit, is mostly out of plane.
Interestingly, as already stated, this polarization can be flipped simply by inverting 
the current.

The magnitude of the effect strongly depends on the strength of the spin orbit coupling. 
A quick order of magnitude estimate can be obtained by inspecting the caption of our 
Figure \ref{sH} while making reference to our model hamiltonian 
(Eq.~(\ref{Hzerobound})).

While the problem can be completely solved by numerical means we will show that a careful 
study of the perturbation expansion in the spin orbit coupling constant is crucial in
understanding the interplay of the various components of the polarization.

The paper is organized as follows: Section \ref{formulation} defines our model and details
our formulation; Section \ref{solution} describes the numerical and perturbative solutions 
to the single particle problem providing a discussion of various properties of the 
eigenfunctions and eigenenergies. Section \ref{edgeM} is devoted to the study of the total
spin polarization in the presence of a steady state current. Many of our most relevant results 
are presented in this Section. Finally Section \ref{discussion} provides a discussion of the
most interesting physics including that of the crossover of the in plane component of
the polarization and our conclusions.
\section{Formulation of the problem and boundary conditions}
\label{formulation}
The model Hamiltonian we consider is the following:
\begin{equation}
\label{Hzerobound}
\hat{H}=\frac{\hat{\mathbf{p}}^2}{2m}+ \alpha(\hat\sigma_x \hat p_y -
\hat\sigma_y \hat p_x)+V(\hat y)  ~,
\end{equation}
where the motion is restricted to the $x,y$ plane and the lateral confining potential 
is given by:
\begin{eqnarray}
\label{potential}
&&V(y)=
\left\{
\begin{array}{cl}
0 \qquad &   \mathrm{for\,\, } |y|<W/2 \\
\infty \qquad & \mathrm{for \,\,} |y| \geq W/2 ~.
\end{array}
\right.
\end{eqnarray} 
Since the system is translationally invariant along the $x$ direction, the eigenfunctions 
of the problem can be constructed by linear superposition of free space ($V(y)=0$) 
eigenstates with definite values of $k_x$. Furthermore the reflection symmetry 
$x \to -x$ allows one to generally assume $k_x>0$.

\begin{figure}
\begin{center}
\includegraphics[width=0.49\textwidth]{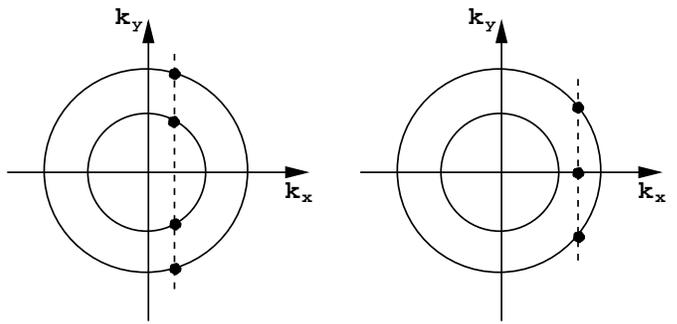}
\caption{\label{4states} Possible free space states that can be superimposed
to obtain confined eigenstates. On right we show a situation for which the
two states of the upper branch have ${\rm Re}[{k_y}]=0$.}
\end{center}
\end{figure}

For reference, we recall here that the free space electron eigenfunctions and 
eigenenergies are given by
\begin{equation}
\label{phi0kpm}
\varphi_{ {\bf k} , \pm}^{(0)} ( x, y ) ~=~
\frac{e^{i (k_x x + k_y y )} }{\sqrt{2 L^2}}
\left(
\begin{array}{c}
\pm 1\\
i e^{i\phi_{\bf k}}
\end{array}
\right) ~,
\end{equation}
\begin{equation}
\label{E0kpm}
\epsilon_{\mathbf{k}\pm } = 
\frac{\hbar^2 (k_x^2 + k_y^2) }{2 m}\mp \alpha \hbar \sqrt{k_x^2 + k_y^2} ~,
\end{equation}
where $L$ is the linear size of the system and $\phi_{\bf k}$ is the angle 
between the direction of the wave vector and the $x$-axis. Interestingly the
Rashba spin-orbit coupling forces each plane wave state to have its own distinct
orientation of the spin quantization axis. This direction lies in the $x-y$ plane 
and makes an angle of $\frac{\pi}{2}$ with ${\bf k}$. 
These states form two split bands which are characterized by opposite chirality in
the sense that states with the same wave vector have opposite spin directions in the
two bands.\cite{bychkov84a, bychkov84b}
Moreover, because of time reversal, in the presence of Rashba spin orbit, states 
within the same band with opposite wave vectors have opposite spins.
The corresponding surfaces (lines) at constant energy $\epsilon$ are concentric 
circles with radii $K_\pm$. These are given as the positive solutions of 
the equation
\begin{equation}
\label{abskpm}
\epsilon=\frac{\hbar^2 K_\pm^2}{2 m}\, \mp \,\hbar\, \alpha \,K_\pm ~.
\end{equation}
Constant energy lines are schematically displayed in Figure \ref{4states}.

At a given energy $\epsilon$, there are only four free space eigenstates that can be 
combined to form solutions satisfying the zero boundary condition of the inhomogeneous 
problem. This is depicted in Figure \ref{4states}: for each spin subband $\pm$ we have 
two states with opposite value of $k_y$. 
The four possible values of $k_y$ are $\pm k^\pm_y$, 
with 
\begin{equation}
\label{defkypm}
k^\pm_y=\sqrt{K_\pm^2-k_x^2}  ~,
\end{equation}
where $K_\pm$ are given in Eq.~(\ref{abskpm}) above.

Since we are interested in spatially undamped solutions, we always take $k_x<K_+$. 
However, it will necessary to include the case $K_-<k_x<K_+$, where
$k_y^-=i |k_y^-|$ is purely imaginary. We also define the angles $\phi_\pm$ by 
means of:
\begin{equation}
\label{defangles}
\phi_\pm=\arctan{\frac{k^\pm_y}{k_x}} ~.
\end{equation}
If $k_y^-=i |k_y^-|$ we have that also $\phi_-$ is purely imaginary and from the 
above formula we get $\phi_-=i \operatorname{arctanh}{\frac{|k^-_y|}{k_x}}$.

It is convenient to consider the following two couples of (unnormalized)
states, which are also eigenstates of the reflection symmetry $y \to -y$:
\begin{eqnarray}
\label{states}
\psi_{\pm\pm}(\epsilon,k_x,y)=
\left(
\begin{array}{cc}
& \cos{( k^\pm_y y-\frac{\phi_\pm}{2})} \\
\pm i & \cos{( k^\pm_y y+\frac{\phi_\pm}{2})}
\end{array}
\right)  e^{i k_x x} ~,  \\
\psi_{\pm\mp}(\epsilon,k_x,y)=
\left(
\begin{array}{cc}
& \sin{( k^\pm_y y-\frac{\phi_\pm}{2})} \\
\pm i & \sin{( k^\pm_y y+\frac{\phi_\pm}{2})}
\end{array}
\right)   e^{i k_x x}  ~.
\end{eqnarray}
The first $\pm$ subscript refers to the chirality of the
spin subband while the second $\pm$ subscript refers to the 
reflection operation $y \to -y$. The eigenstates in the confined system are 
then expressed in the following way:
\begin{eqnarray}
\label{confinedststes1}
&\varphi^+_n(k_x,y)=c_n^{++}\psi_{++}+c_n^{-+}\psi_{-+}  ~,\\
\label{confinedststes2}
&\varphi^-_n(k_x,y)=c_n^{+-}\psi_{+-}+c_n^{--}\psi_{--}   ~,
\end{eqnarray}
where the superscript in $\varphi_n^\pm$ refers to the reflection parity, and 
not to the chirality. As one would expect, the solutions of the confined system 
are an admixture of states with different chirality. A spin subband index will 
be introduced as an approximate quantum number in the case of small spin orbit 
in a following Section where we present a perturbative treatment of the problem. 

Imposing the boundary condition at $y=W/2$ for the $\varphi^\pm_n$ leads to the 
following conditions:
\setlength\arraycolsep{0pt}
\begin{eqnarray}
\label{boundcond1}
\left(
\begin{array}{ccc}
\cos{(\frac{k^+_y  W-\phi_+}{2})} &~& \sin{( \frac{k^-_y  W-\phi_-}{2})}  \\
i  \cos{(\frac{k^+_y  W+\phi_+}{2})} &~& -i  \sin{( \frac{k^-_y  W+\phi_-}{2})}
\end{array}
\right)
\left(
\begin{array}{c}
c_n^{++} \\
c_n^{-+}
\end{array}
\right)=0   ~,\qquad \\
\label{boundcond2}
\left(
\begin{array}{ccc}
  \sin{(\frac{k^+_y  W-\phi_+}{2})} &~& \cos{( \frac{k^-_y  W-\phi_-}{2})}  \\
i  \sin{(\frac{k^+_y  W+\phi_+}{2})} &~& -i  \cos{( \frac{k^-_y  W+\phi_-}{2})}
\end{array}
\right)
\left(
\begin{array}{c}
c_n^{+-} \\
c_n^{--}
\end{array}
\right)=0  ~. \qquad
\end{eqnarray}
At fixed $k_x$, the the determinants of (\ref{boundcond1}) and (\ref{boundcond2})
are functions of $\epsilon$, through the implicit dependence of $k^\pm_y$ and $\phi_\pm$.
The zeros of these determinants provide two sets of discrete energies $\epsilon^\pm_n(k_x)$,
corresponding to states of opposite parity. 



\begin{figure}
\begin{center}
\includegraphics[width=0.49\textwidth]{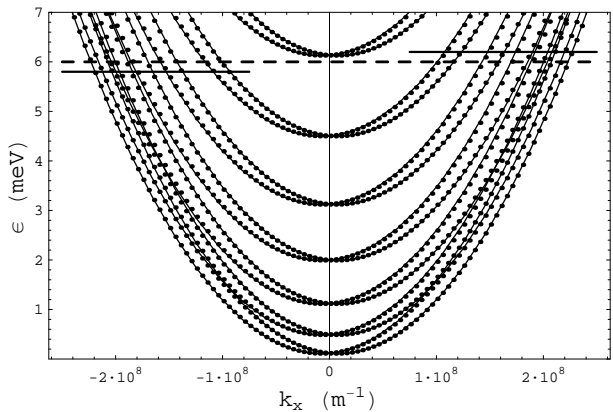}
\caption{\label{spectrum} Energy spectrum of a $W=0.1~\mu{\rm m}$ wide quantum wire. 
For clarity the exact energy bands are only shown (dots) for discrete values of $k_x$, 
while the perturbative results from Eq.~(\ref{energyexp}) are also plotted (solid lines) 
for comparison. 
The parameters values are as follows: 
$\hbar\alpha=2~10^{-9} {\rm meV}\,{\rm m}$ and  $m=0.3 m_0$. 
The horizontal dashed line marks a Fermi level of $6~{\rm meV}$ and is significant 
in relation to Figures \ref{spindensitiesZ}, \ref{spindensitiesY}, and \ref{sH}. 
The corresponding electron density is $0.67~10^{12}~{\rm cm}^{-2}$.}
\end{center}
\end{figure}
\begin{figure}
\begin{center}
\includegraphics[width=0.49\textwidth]{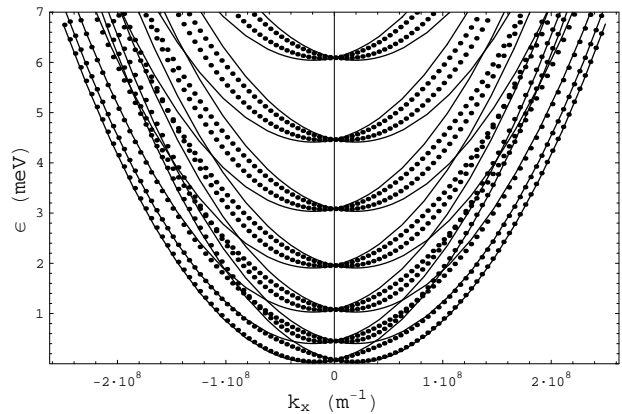}
\caption{\label{spectrum2} Same as in Figure \ref{spectrum} but with a larger
value of the spin orbit coupling: $\hbar\alpha=5~10^{-9}~{\rm meV}\,{\rm m}$. 
Strong deviations from perturbation theory are due to anti-crossing of bands.}
\end{center}
\end{figure}

If one is not interested in separating states with different reflection parity, 
one can combine the conditions ensuing from (\ref{boundcond1}) and (\ref{boundcond2})
into the compact equation:
\begin{eqnarray}
\label{spectrumpm}
&&(1-\cos{k_y^+W}\cos{k_y^-W})\sin{\phi_+}\sin{\phi_-}\\
&&+(1-\cos{\phi_+}\cos{\phi_-})\sin{k_y^+W}\sin{k_y^-W}=0 
~, \nonumber
\end{eqnarray} 
which gives the whole spectrum for a given $k_x$. 
The eigenfunctions, dropping the reflection parity index, are henceforth quite 
generally denoted as $\varphi_\nu (k_x,y)$.

We finally notice that for values of $\epsilon$ and $k_x$ such that 
$k^-_y$ is purely imaginary, the structure of the above formulas still holds, 
and one need only substitute hyperbolic for trigonometric functions.

\section{Solution of the problem}
\label{solution}
\subsection{Exact solution at $k_x=0$}
A case in which the solutions can be expressed in simple closed form is that of
$k_x=0$.  For this particular value of $k_x$ it is immediate to see that
Eq.~(\ref{Hzerobound}) admits solutions which are eigenstates of $\hat\sigma_x$. 
Assuming the following form for the wavefunctions 
\begin{equation} 
\label{solzerokstate} 
e^{\mp i \frac{m \alpha}{\hbar} y} \varPhi(y)|\pm \rangle_x ~, 
\end{equation} 
$\varPhi(y)$ is found to satisfy:
\begin{equation}
\label{simplschroe}
\left[
-\frac{\hbar^2}{2m}\frac{\partial^2}{\partial y^2}
+ V(y) -\frac{1}{2} m \alpha^2\right]\varPhi(y)=
\epsilon \, \varPhi(y) ~.
\end{equation}
This result is quite general holding for an arbitrary form of the confining potential 
$V(y)$. For the case of an infinite well we obtain:
\begin{eqnarray}
\label{finalsol}
\epsilon_{n} = \frac{\hbar^2}{2m}\left(\frac{\pi n}{W} \right)^2-\frac{1}{2} 
m \alpha^2~, 
\end{eqnarray} 
where $n$ is a positive integer. 
For every eigenvalue there is an additional twofold spin degeneracy.
\subsection{Numerical results}

\begin{figure*}
\begin{center}
\hspace{1cm}\includegraphics[width=0.7\textwidth]{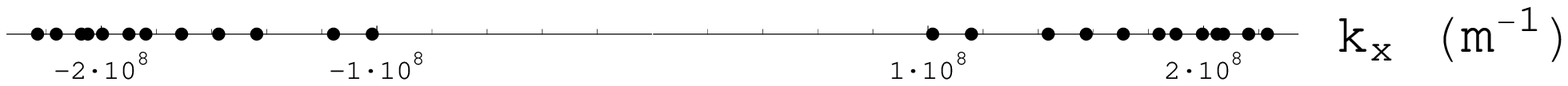}
\includegraphics[width=0.6\textwidth]{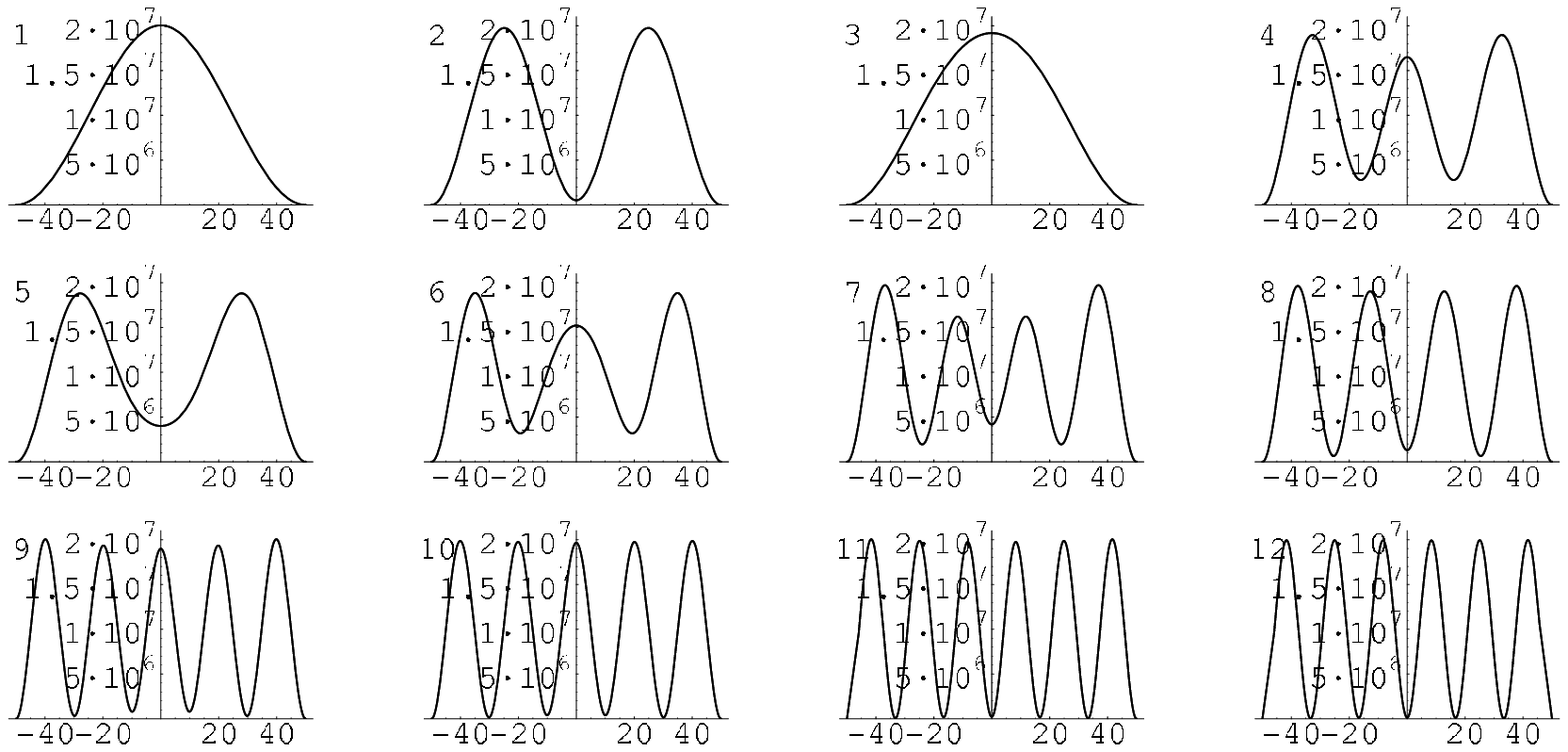}
\caption{\label{densities} 
Number density $\mathcal{N}_\nu(y)$ (in ${\rm m}^{-1}$ units) 
of individual eigenstates at the Fermi level of a quantum wire. The parameters
are the same as in Figure \ref{spectrum}. The wave vector values are determined by the
crossings of the bands with the horizontal dashed line at $6~{\rm meV}$ in 
Figure \ref{spectrum} and are given at the top (dots on the $k_x$ line). 
The corresponding $\mathcal{N}_\nu(y)$ functions are shown in the panels,
where the numbers label the first twelve wave vectors, counted from left to right 
on the top $k_x$ line. States with opposite wave vectors have the same density.}
\end{center}
\end{figure*}

\begin{figure*}
\begin{center}
\hspace{1cm}\includegraphics[width=0.7\textwidth]{wavevectors.eps}
\includegraphics[width=0.6\textwidth]{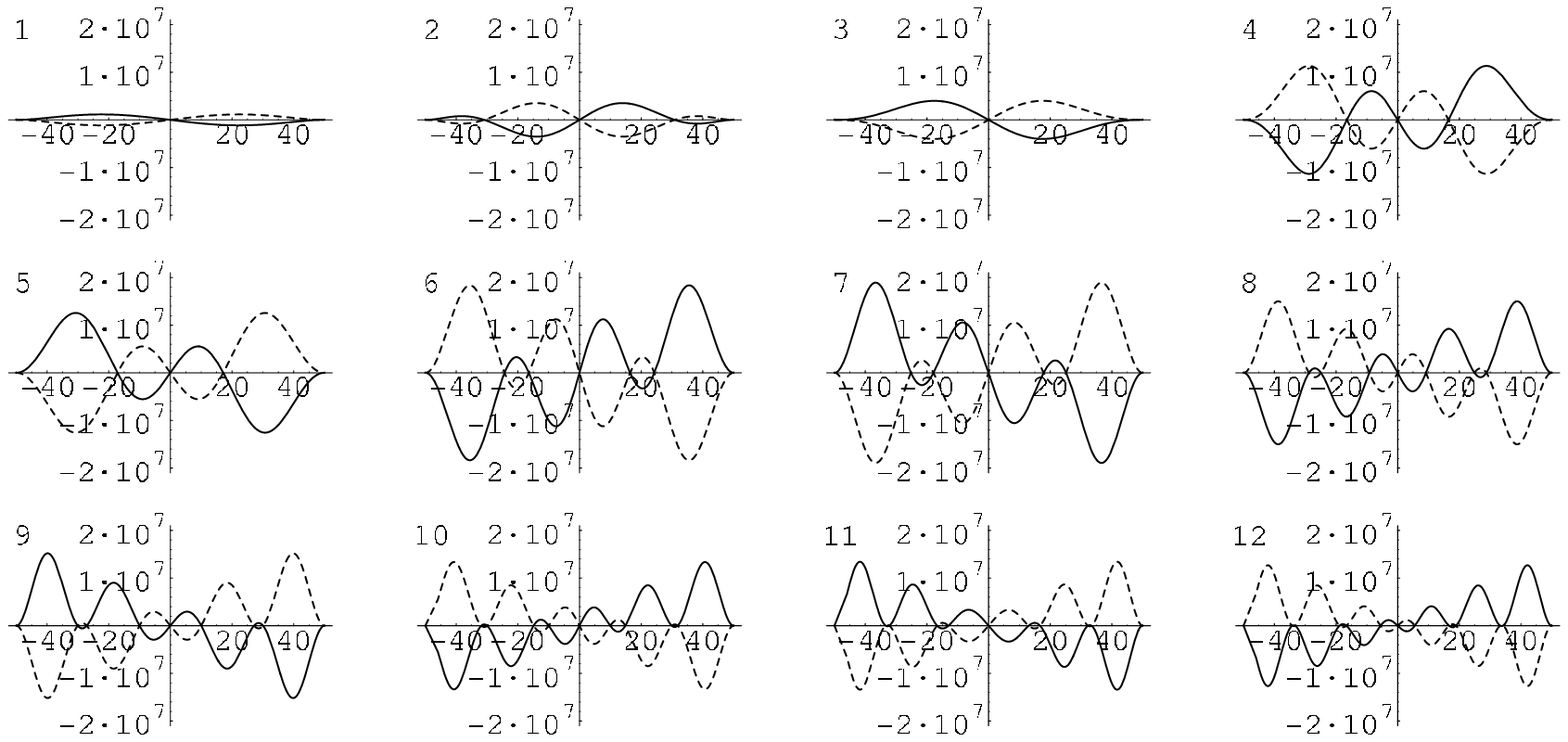}
\caption{\label{spindensitiesZ} Spin polarization density $\mathcal{P}_\nu^z(y)$ 
along the $z$ direction (in ${\rm m}^{-1}$ units) of the same eigenstates 
of Figure \ref{densities}. States with opposite wave vectors have 
polarization of opposite sign. Solid lines correspond to the first twelve $k_x$ values, 
counted from left to right on the top $k_x$ line, while dashed lines refer to $-k_x$.}
\end{center}
\end{figure*}

\begin{figure*}
\begin{center}
\hspace{1cm}\includegraphics[width=0.7\textwidth]{wavevectors.eps}
\includegraphics[width=0.6\textwidth]{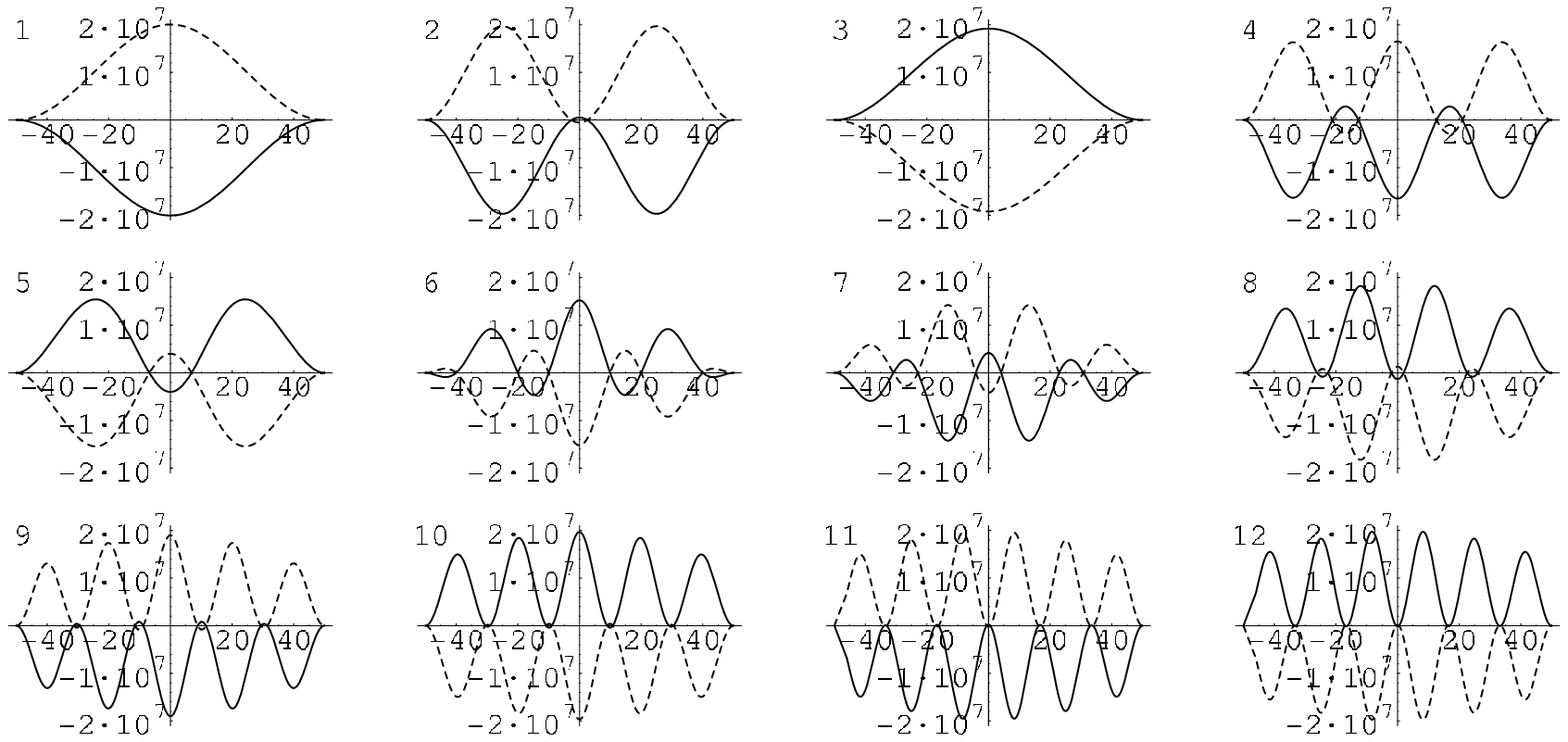}
\caption{\label{spindensitiesY}  Spin polarization density $\mathcal{P}_\nu^y(y)$ 
along the $y$ direction (in ${\rm m}^{-1}$ units) of the same eigenstates 
of Figure \ref{densities}. States with opposite wave vectors have 
polarization of opposite sign. Solid lines correspond to the first twelve $k_x$ values, 
counted from left to right on the top $k_x$ line, while dashed lines refer to $-k_x$.}
\end{center}
\end{figure*}

We reproduce in Figures \ref{spectrum}-\ref{spindensitiesY} numerical results, 
calculated with the methods of the previous Section. 
In Figure \ref{spectrum} and \ref{spectrum2} we present two examples of the energy 
spectrum as function of $k_x$ for different values of the spin-orbit coupling $\alpha$. 

The properties of the eigenfunctions are perhaps more interesting and can be best
exemplified by the corresponding number and spin polarization densities.
Figures \ref{densities}, \ref{spindensitiesZ} and \ref{spindensitiesY} refer to these properties of the 
eigenfunctions, and have been obtained with the same parameters of Figure \ref{spectrum} 
and an occupation characterized by a Fermi energy $\epsilon_F=6~{\rm meV}$.
In particular, we show there for each state the relative number density 
$\mathcal{N}_\nu(k_x,y)$ and spin polarization density $\vec{\mathcal{P}}_\nu(k_x,y)$. 
These quantities are defined as follows:  
\begin{equation}
\label{Ndef}
\mathcal{N}_\nu(k_x,y) ~ =~ \langle\varphi_\nu(k_x,y')|\delta(y-\hat y')
|\varphi_\nu(k_x,y')\rangle ~,
\end{equation}
\begin{equation}
\label{Pdef}
\vec{\mathcal{P}}_\nu(k_x,y) ~=~ 
\langle\varphi_\nu(k_x,y')|\hat{\vec\sigma} \delta(y-\hat y')|\varphi_\nu(k_x,y')\rangle ~.
\end{equation}

From simple symmetry considerations, it follows that the polarization is vanishing in the
$x$ direction, along the wire, and states with the same energy and 
opposite values of $k_x$ have the same density but opposite spin polarization. 
As it will be discussed in a following Section, this fact allows in principle to produce 
a net spin polarization along the $y$ and $z$ direction, by driving an electrical current along 
the wire.


\subsection{Perturbation theory}
Although the numerical analysis provides the complete answer to the problem, 
useful insight can be gained from an examination of the perturbative approach. 
We therefore consider next the case of small spin orbit coupling and will assume 
that the Rashba term
can indeed be treated as a perturbation. 
In order to simplify the notation, in this Section we consider a potential that 
confines the electrons in $0<y<W$ rather than $-W/2<y<W/2$. 
For simplicity we will label the perturbed eigenstates via the (approximate) quantum 
numbers of the unperturbed solutions. Accordingly we write
\begin{equation}
\label{pertstate}
\varphi_{n\pm}(k_x,y)=\varphi^0_{n\pm}(k_x,y)+O(\alpha) ~,
\end{equation}
where
\begin{equation}
\label{unpertstate}
\varphi^0_{n\pm}(k_x,y)=
\sqrt{\frac{2}{L W}}\, e^{i k_x x}\sin{\frac{n \pi y}{W}}
\, | \pm \rangle_y   ~, 
\end{equation}
with $n$ a positive integer.
We also find convenient to choose the spinors along $y$, in such a way that the states are
eigenstates of the $y$ reflection symmetry. The relevant matrix elements of the spin 
orbit interaction are:
\begin{eqnarray}
\label{matrixelements}
\langle \varphi^0_{n'\pm}|\hat R_0|\varphi^0_{n\pm}\rangle &=&
\mp \hbar\alpha k_x \delta_{n n'}\delta_{k_x k_x'}  ~,\\
\langle \varphi^0_{n'\mp}|\hat R_0|\varphi^0_{n\pm}\rangle &=&
\mp \frac{2 \hbar\alpha}{W}\frac {n n'(1-(-1)^{n+n'})}{(n^2-n'^2)}\, \delta_{k_x k_x'}\nonumber ~.
\end{eqnarray}

We can easily obtain the form of the perturbation expansion, up to
second order:
\begin{eqnarray}
\label{energyexp}
\epsilon_{n\pm}(k_x)&=&\frac{\hbar^2}{2m}\left(\frac{\pi n}{W^2}\right)^2 
+\frac{\hbar^2k_x^2}{2m} \nonumber\\
&&\mp \hbar \alpha k_x -\frac{1}{2}m \, \alpha^2+O(\alpha^3)  ~. \quad
\end{eqnarray}
Higher order terms are also readily obtained.

The necessary condition for the validity of the perturbative expansion is that 
the matrix elements (\ref{matrixelements}) be smaller than the unperturbed 
energy differences. The latter are of order $\frac{\hbar^2}{m W^2}$. This gives:
\begin{equation}
\label{pertcondns}
\alpha \ll \frac{\hbar}{m k_x W^2} \quad {\rm and} \quad   \alpha \ll \frac{\hbar}{mW}~.
\end{equation}
Clearly the perturbative expansion fails for large values of $k_x$, or for a 
channel with large $W$. 



The perturbation theory allows us to explicitly calculate a number of interesting
quantities. In particular the $z$-component of the spin polarization density of the 
confined states (see Eq.~(\ref{Pdef})) is given by:
\begin{eqnarray}
\label{spindensityn}
\mathcal{P}^z_{n\pm}(k_x,y)=\langle \varphi_{n\pm}(k_x,y') |\hat \sigma_z \delta(y-\hat y')|  
\varphi_{n\pm}(k_x,y') \rangle \qquad \\
=\mp\frac{\alpha m}{\hbar} F_n(y/W)-\left(\frac{\alpha m W}{\hbar}\right)^2 k_x G_n(y/W) 
+O(\alpha^3) ~,\nonumber 
\end{eqnarray}
where the functions $F_n$ and $G_n$ are given by:
\begin{eqnarray}
\label{defFG}
F_n(y/W) &=& \frac{64}{\pi^2} {\sum_{m}}^\prime \frac{n m}{(n^2-m^2)^2} 
\sin{\frac{\pi n y}{W}}\sin{\frac{\pi m y}{W}} ~,\nonumber\\
G_n(y/W)&=&\frac{256}{\pi^4} {\sum_m}^\prime \frac{n m}{(n^2-m^2)^3} 
\sin{\frac{\pi n y}{W}}\sin{\frac{\pi m y}{W}}  ~, \qquad
\end{eqnarray} 
where the sum extends over positive even (odd) integers $m$ for odd (even) $n$.

In the same way, while $\mathcal{P}^x_{n\pm}(k_x,y)$ vanishes, the corresponding
perturbative result for the polarization along $y$ of the $n \pm$ states can
be expressed as:
\begin{eqnarray}
\label{spindensitynY}
\mathcal{P}^y_{n\pm}(k_x,y)=\langle \varphi_{n\pm}(k_x,y') |\hat \sigma_y \delta(y-\hat y')|
\varphi_{n\pm}(k_x,y') \rangle \qquad \\
=\pm \frac{2}{W}\sin^2{\frac{\pi n y}{W}}\mp\left(\frac{\alpha m}{\hbar}\right)^2 W H_n(y/W)
+O(\alpha^3) ~,\nonumber
\end{eqnarray}
where the function $H_n$ is defined as follows:
\begin{eqnarray}
\label{defH}
&&H_n(y/W)=\frac{128}{\pi^4}\Bigg[
{\sum_{m}}^\prime\frac{m^2 n^2}{(n^2-m^2)^4}
\sin^2{\frac{\pi n y}{W}}
\nonumber \\
&&+2{\sum_{m,n'}}^\prime \frac{m^2 n n'}{(n^2-m^2)^2(n^2-n'^2)(m^2-n'^2)}
\sin{\frac{\pi n y}{W}}\sin{\frac{\pi n' y}{W}} \nonumber \\
&&+{\sum_{m,m'}}^\prime\frac{m m'n^2}{(n^2-m^2)^2(n^2-m'^2)^2}
\sin{\frac{\pi m y}{W}}\sin{\frac{\pi m' y}{W}} \Bigg] ~,
\end{eqnarray}
where the sums extend over the positive integers. In particular while $n'$ is of the same 
parity as $n$, both $m$ and $m'$ are of opposite parity. Moreover $n' \neq n$.

Notice that as expected, in all these formulas the substitution $k_x\to -k_x$ and 
$\pm\to\mp$ gives a state with the same energy and opposite spin polarization.

\section{Edge spin polarization}
\label{edgeM}

We consider here the spin accumulation at the boundaries of the wire
in the presence of an electric current in the $x$ direction.  

Although the response of a system of electrons subject to Rashba spin orbit and
an applied electric field is quite complicated and requires a careful 
analysis,\cite{aronov89,aronov91} for simplicity sake we will model here a current 
carrying steady state by assuming a modified electron occupation distribution. 
In particular we will assume that a finite chemical potential difference 
$\delta\mu=\mu_+ -\mu_-$ is established between the right moving and left moving 
states. This is depicted in Figure \ref{spectrum} by the two short horizontal 
solid lines (whose separation from the equilibrium Fermi level - dashed - is not in scale). 
The present model is meant to describe ballistic transport with all the chemical 
potential drop occurring at the contacts at $x =\pm \infty$.

Within this framework, the total spin polarization density due to the current 
flux can be obtained in the linear regime from the relation:
\begin{equation}
\label{sHballistic}
\vec{\mathcal{M}}(y) ~=~ \pm \frac{g \mu_B}{2}
\tilde{\sum_{\nu}} \frac{ \delta\mu }{ 2 \pi \hbar \, v_{F\nu} } 
\vec{\mathcal{P}}_{\nu}(k_{F\nu},y) ~,
\end{equation} 
where the sum is restricted to the occupied one-dimensional subbands
with $k_{F\nu}$ and $v_{F\nu}$ being the (positive) Fermi wave vector 
and Fermi velocity relative to the occupied band as labeled by the index $\nu$. 
In this formula the positive sign refers to the case of hole transport. As a
consequence the direction of the magnetization only depends on the current 
direction and not on the type of carrier involved.

\begin{figure}
\begin{center}
\includegraphics[width=0.4\textwidth]{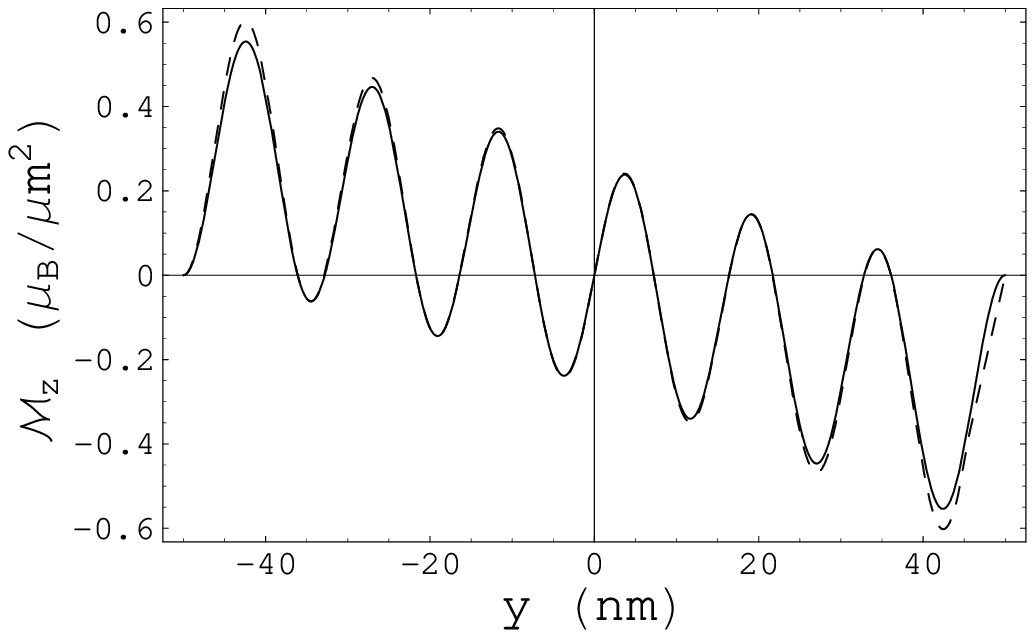}\\
\includegraphics[width=0.4\textwidth]{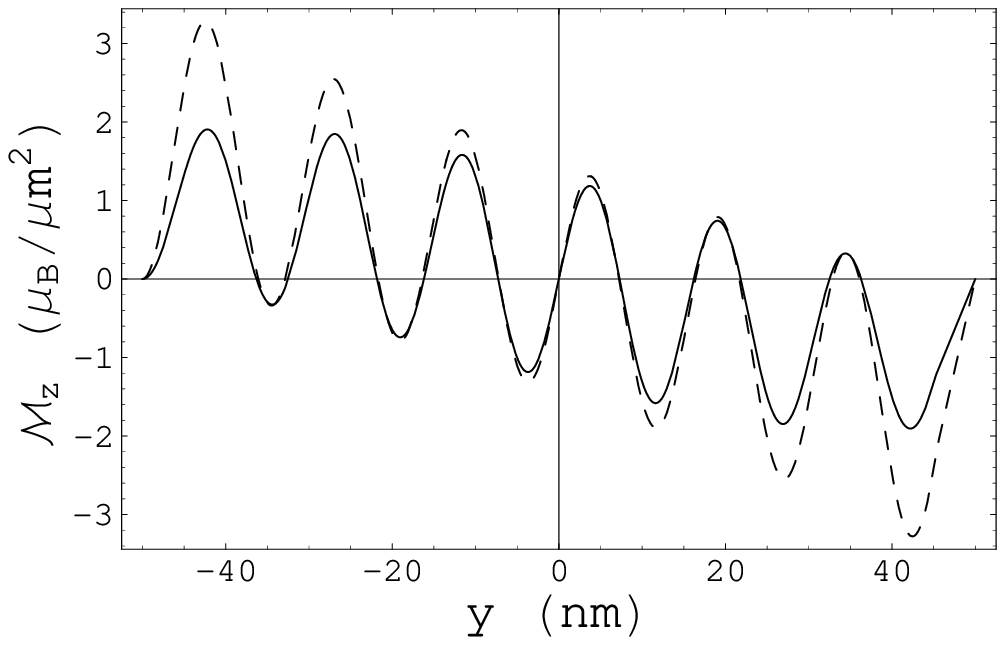}
\caption{\label{sH} Spin polarization density in the $z$ direction 
of a $W=0.1~\mu{\rm m}$ wide quantum wire, 
for carriers with effective mass $m=0.3 m_0$ and $g=2$. 
The Fermi energy is $\epsilon_F=6~{\rm meV}$, which corresponds
to an electron density of $0.67~10^{12}~{\rm cm}^{-2}$. 
We used $\delta\mu=0.1~{\rm meV}$ which gives 
a current in the wire of $23~{\rm nA}$.
In the top panel $\hbar\alpha=1.5~10^{-9}~{\rm meV}\,{\rm m}$ while in the bottom
one $\hbar\alpha=3.5~10^{-9}~{\rm meV}\,{\rm m}$. The perturbative result is 
also plotted (dashed). 
As a comparison of the magnitude of the effect, the Pauli susceptibility is 
$\chi_P=72.5~\mu_B/(\mu{\rm m}^{2} \, {\rm T})$.}
\end{center}
\end{figure}

The perturbative result for the $z$ component $\mathcal{M}_z$ is readily obtained 
from Eqs.~(\ref{defFG}):
\begin{equation}
\label{perturbSH}
\mathcal{M}_z(y) ~\simeq~ \mp  \delta \mu 
\frac{m g \mu_B }{2\pi \hbar^2}\left(\frac{m\alpha W}{\hbar}
\right)^2 \tilde{\sum_n} G_n(y/W) ~,
\end{equation} 
where the sum is limited to the positive integers corresponding to the occupied bands.
The results of the (exact) numerical calculations of $\mathcal{M}_z(y)$ are shown 
in Figure \ref{sH}. 

\begin{figure}
\begin{center}
\includegraphics[width=0.4\textwidth]{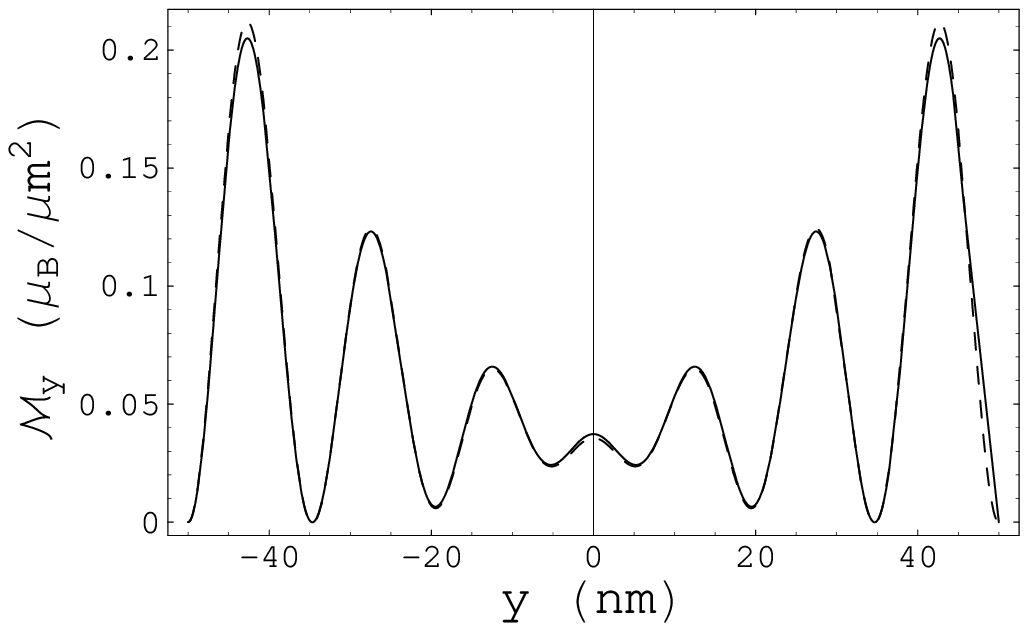}\\
\includegraphics[width=0.4\textwidth]{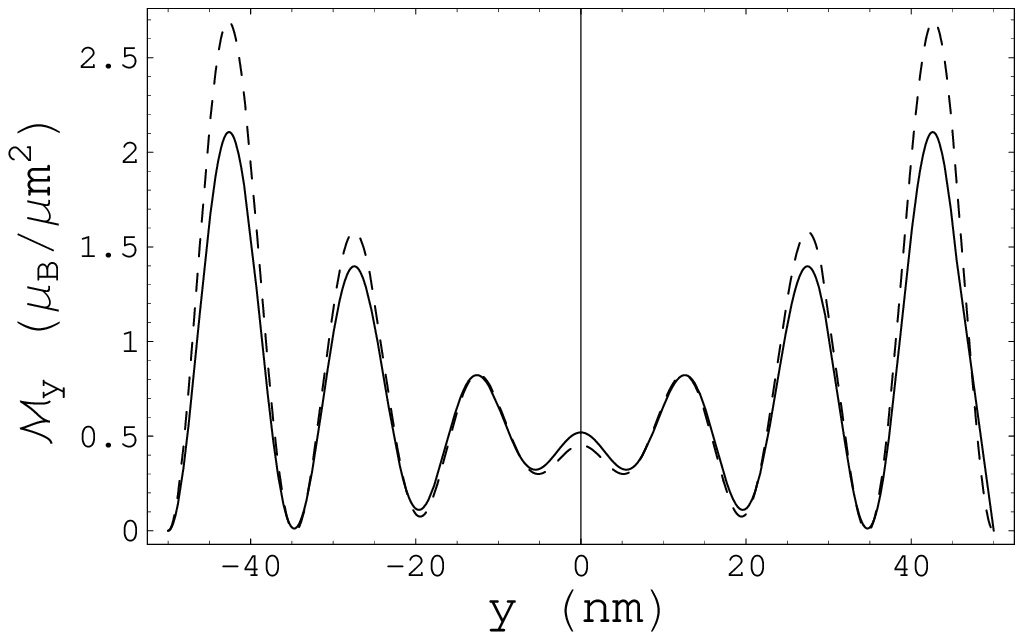}
\caption{\label{sHy} Spin polarization in the $y$ direction corresponding
to the same parameters of the previous Figure \ref{sH}.
In the top panel $\hbar\alpha=1.5~10^{-9}~{\rm meV}\,{\rm m}$ while in the bottom
one $\hbar\alpha=3.5~10^{-9}~{\rm meV}\,{\rm m}$. The leading perturbative 
contribution (cubic in $\alpha$) is  also plotted (dashed).}
\end{center}
\end{figure}

Interestingly the expression given in (\ref{spindensitynY}) is not sufficient to obtain the
leading corresponding perturbative expression for $\mathcal{M}_y(y)$. The reason is simply 
that in the confined geometry the first finite contribution to this quantity is clearly 
of order $\alpha^3$ (see also below). 
This term can be of course readily calculated. The results for the exact calculation 
of $\mathcal{M}_y(y)$ are plotted in Figure \ref{sHy} alongside the perturbative 
expression.

It is remarkable that the $z$ component of the magnetization is larger for small 
values of the spin-orbit coupling constant while the two components become comparable 
for larger values.  Furthermore, while $\mathcal{M}_z(y)$ changes sign and displays 
an asymmetric behavior across the wire, $\mathcal{M}_y(y)$ remains of the same sign 
and is instead symmetric.

As expected, in all cases, there is good agreement with the approximate 
analytical formula (\ref{perturbSH}) for small values of $\alpha$. 


\begin{figure}
\begin{center}
\includegraphics[width=0.4\textwidth]{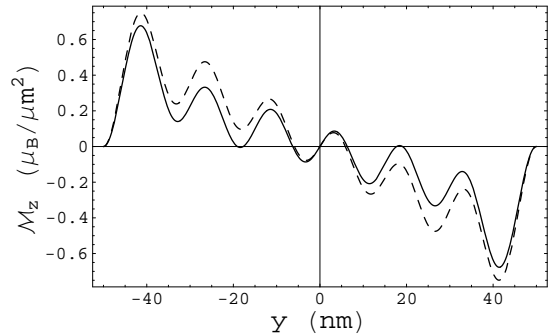}
\caption{\label{sHtau} Magnetization density of a $W=0.1~\mu{\rm m}$ wide quantum wire, 
for carriers with same parameters as the first panel of Figure \ref{sH}. 
We used $E=1~{\rm kV/m}$ and a constant value for $\tau=7.3~10^{-13}~{\rm s}$, 
which gives the same current in the wire of Figure \ref{sH}. 
The perturbative result is also plotted (dashed). The magnetization is somewhat 
different from the previous case, where the oscillations of the magnetization 
are more pronounced. The overall sign of the effect is the same.}
\end{center}
\end{figure}

Our theory can be extended to the case of non ballistic transport. A simple treatment
of such a situation can be obtained by setting
$\delta\mu \to e E v_{F \nu} \tau_{\nu}$ in Eq.~(\ref{sHballistic}).
Here $E$ is the magnitude of the driving electric field
and $v_{F \nu} \tau_{\nu}$ is the elastic mean free path appropriate for the states of
the one dimensional subband $\nu$, a length scale that we assume much larger than the 
width $W$ of the wire. 
By taking for simplicity the same scattering time for all subbands 
$\tau_{\nu} = \tau$ the ensuing magnetization can be obtained. The results 
for $\mathcal{M}_z(y)$ are plotted in Figure \ref{sHtau}. We notice that 
the oscillations of the magnetization are somewhat damped as compared to the 
ballistic case. On the other hand, the overall sign of the effect is unchanged, 
the behavior being in qualitative agreement with the experimental results 
of Ref.~\onlinecite{kato04b}. We haste to state however that our model is rather different from 
the situation of a dirty three-dimensional sample dealt with in the experiment. 
\section{Discussion}
\label{discussion}
\begin{figure}
\begin{center}
\includegraphics[width=0.25\textwidth]{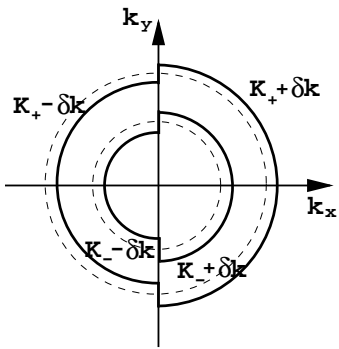}
\caption{\label{distorted_occupation}
Distorted momentum space occupation distribution used to calculate the 
bulk value of the magnetization. $K_\pm$ are defined in Eq.~(\ref{abskpm}).}
\end{center}
\end{figure}

As we have noted, while for small spin orbit the out of plane (z-axis) polarization
$\mathcal{M}_z(y)$ induced by an x-axis current 
is quadratic in the coupling constant, the in plane (y-axis) polarization 
$\mathcal{M}_y(y)$ is much smaller, first appearing in third order. 
This result is only in apparent contradiction with the well known corresponding 
bulk result in two dimensions which entails a linear dependence of $\mathcal{M}_y(y)$
on the coupling constant. 
Within the simplified model used in Section \ref{edgeM} such a behavior can be 
readily derived by assuming the distorted momentum space electron distribution function
of Figure \ref{distorted_occupation}. The final expression is given by
\begin{equation}
\label{linar_inplane}
\frac{1}{L^2}  \langle \hat \sigma_y \rangle ~=~ 
\frac{\alpha \,  m^{\frac32} \, \delta\mu}
{\pi^2 \hbar^2 \sqrt{2 \epsilon_F+ m \alpha^2}} ~,
\end{equation}
where $\delta\mu = 2 \hbar v_F \delta k $.
As expected for small $\alpha$ $\mathcal{M}_y(y)$ is linear in the coupling 
constant.\cite{commQwire1}

The difference lies with a quenching of the y-axis polarization due to the 
confining potential. As it can be surmised by inspecting Eq.~(\ref{spindensitynY}),
this effect can be readily seen to originate from the lowest order cancellation 
of the contributions to $\mathcal{M}_y(y)$ stemming from states 
within the same spin-split one dimensional band. The cancellation only occurs when 
the spin-splitting is smaller than the energy quantization associated with the 
one-dimensional confining potential. This is of course not operational in the bulk since
in such a situation the energy spectrum is continuum.

\begin{figure}
\begin{center}
\includegraphics[width=0.4\textwidth]{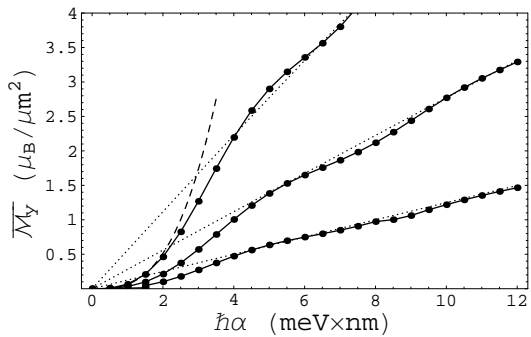}
\caption{\label{crossover} 
Average spin polarization density in the $y$ direction as a function of 
$\alpha$. The parameters used are: $W=0.15~\mu{\rm m}$, $\epsilon_F=2.5, 10, 40~{\rm meV}$,
$\delta \mu=0.05~{\rm meV}$, and $m=0.3 m_0$ (mass as in previous Figures). Here
the dots represent the (exact) numerical result, while the dotted curves are
obtained from Eq.~(\ref{linar_inplane}) in the text. The dashed line is a guide to the eye
representing a pure cubic behavior.}
\end{center}
\end{figure}

The crossover between these two regimes approximately occurs when the spin-splitting 
equals the energy spacing between the eigenvalues of the confining potential along the
$y$ direction, i.e. for $\frac{m \alpha W}{\hbar} \simeq 1$.
In order to exemplify this phenomenon we have plotted in Figure \ref{crossover} the 
spatially averaged magnetization density 
\begin{equation}\label{avgM}
\overline{{\mathcal M}_y} ~=~ \frac{1}{W}\int_{-W/2}^{W/2}
\mathcal{M}_y(y) \mathrm{d}y ~,
\end{equation}
as a function of the value $\hbar\alpha$ of the spin orbit coupling constant.
With the parameters chosen in Figure \ref{crossover} the crossover occurs approximately
for $\hbar\alpha=1.7 \,10^{-9}\,{\rm meV}\,{\rm m}$. The dashed curve superimposed 
to the $\epsilon_F = 2.5~{\rm meV}$ dotted line is purely cubic and provides a useful guide 
to the eye for that case. 

Exactly the opposite mechanism is responsible for the case of $\mathcal{M}_z(y)$,
which, as we have seen, is instead vanishing in the bulk. 
Interestingly, as one can infer from Figure \ref{crossover}, the expression 
of Eq.~(\ref{linar_inplane}), formally valid in the bulk for finite values of $\alpha$ 
when in principle the perturbative condition is inapplicable, still gives good 
results for the case of a rather narrow channel with $W=0.15~\mu{\rm m}$. 

We discuss next how the different behaviors of the two components of the spin 
polarization we have just discussed are compatible with a quite general and rigorous 
argument regarding the parity of $\mathcal{M}_z(y)$ and $\mathcal{M}_y(y)$ with 
respect to the sign of the spin orbit coupling constant. 
Consider the effect on the hamiltonian (including contributions from the electron 
electron interaction and non magnetic impurities) of the application of the operator 
$\hat \sigma_z$. Clearly $\hat \sigma_z \hat H (\alpha) \hat \sigma_z = 
\hat H ( - \alpha) $. Then it immediately follows that given an eigenstate of the 
problem for a given value of $\alpha$, the corresponding solution for $-\alpha$ can 
be obtained simply by application of $\hat \sigma_z$. This in turn implies that, quite
generally, while $\mathcal{M}_z(y)$ is even with respect to $\alpha$, 
$\mathcal{M}_y(y)$ is odd.

We have shown that a net spin polarization is established when a steady state current
is run through the wire obtained by laterally confining a two dimensional electron gas 
in the presence of Rashba spin orbit. Reversing the current inverts the spin 
polarization. The effect is phenomenologically analogous to, yet distinct from 
the spin-Hall effect that has recently been the subject of intense discussion (for 
a useful review, see Ref.~\onlinecite{engel06}) and was demonstrated 
experimentally.\cite{kato04b,wunderlich05,sih05,sih06}
The mechanism relevant to the present discussion simply stems from the structure of 
the electronic wavefunctions in a confined geometry. Accordingly, there is no need 
to make appeal to spin currents or impurity scattering. Moreover its geometric nature 
should leave the phenomenon mostly unchanged in the presence of a moderate amount 
of impurities in the wire. 
A similar effect should also be relevant in the case of electronic states localized
by an impurity potential as well as in the transition to a low density Wigner crystal 
in the presence of Rashba spin orbit.
\begin{acknowledgments}
The authors would like to thank Leonid Rokhinson and 
Yuli Lyanda-Geller for vigorous, useful discussions.
\end{acknowledgments}


\begin{thebibliography}{14}
\expandafter\ifx\csname natexlab\endcsname\relax\def\natexlab#1{#1}\fi
\expandafter\ifx\csname bibnamefont\endcsname\relax
  \def\bibnamefont#1{#1}\fi
\expandafter\ifx\csname bibfnamefont\endcsname\relax
  \def\bibfnamefont#1{#1}\fi
\expandafter\ifx\csname citenamefont\endcsname\relax
  \def\citenamefont#1{#1}\fi
\expandafter\ifx\csname url\endcsname\relax
  \def\url#1{\texttt{#1}}\fi
\expandafter\ifx\csname urlprefix\endcsname\relax\def\urlprefix{URL }\fi
\providecommand{\bibinfo}[2]{#2}
\providecommand{\eprint}[2][]{\url{#2}}

\bibitem[{com({\natexlab{a}})}]{commQwire0}
\bibinfo{note}{That the effects of the electronic interactions can be neglected
  is usually a rather dangerous assumption. A systematic study of the interplay
  of the Coulomb interaction and Rashba spin-orbit in a two dimensional
  electron liquid can be found in S. Chesi and G. F. Giuliani, unpublished.}

\bibitem[{\citenamefont{Aronov and Lyanda-Geller}(1989)}]{aronov89}
\bibinfo{author}{\bibfnamefont{A.~G.} \bibnamefont{Aronov}} \bibnamefont{and}
  \bibinfo{author}{\bibfnamefont{Y.~B.} \bibnamefont{Lyanda-Geller}},
  \bibinfo{journal}{JETP Lett.} \textbf{\bibinfo{volume}{50}},
  \bibinfo{pages}{431} (\bibinfo{year}{1989}).

\bibitem[{\citenamefont{Aronov et~al.}(1991)\citenamefont{Aronov,
  Lyanda-Geller, and Pikus}}]{aronov91}
\bibinfo{author}{\bibfnamefont{A.~G.} \bibnamefont{Aronov}},
  \bibinfo{author}{\bibfnamefont{Y.~B.} \bibnamefont{Lyanda-Geller}},
  \bibnamefont{and} \bibinfo{author}{\bibfnamefont{G.~E.} \bibnamefont{Pikus}},
  \bibinfo{journal}{Sov. Phys. JETP} \textbf{\bibinfo{volume}{73}},
  \bibinfo{pages}{537} (\bibinfo{year}{1991}).

\bibitem[{\citenamefont{Edelstein}(1990)}]{edelstein90}
\bibinfo{author}{\bibfnamefont{V.~M.} \bibnamefont{Edelstein}},
  \bibinfo{journal}{Solid State Commun.} \textbf{\bibinfo{volume}{73}},
  \bibinfo{pages}{233} (\bibinfo{year}{1990}).

\bibitem[{\citenamefont{Kato et~al.}(2004{\natexlab{a}})\citenamefont{Kato,
  Myers, Gossard, and Awschalom}}]{kato04a}
\bibinfo{author}{\bibfnamefont{Y.~K.} \bibnamefont{Kato}},
  \bibinfo{author}{\bibfnamefont{R.~C.} \bibnamefont{Myers}},
  \bibinfo{author}{\bibfnamefont{A.~C.} \bibnamefont{Gossard}},
  \bibnamefont{and} \bibinfo{author}{\bibfnamefont{D.~D.}
  \bibnamefont{Awschalom}}, \bibinfo{journal}{Phys. Rev. Lett.}
  \textbf{\bibinfo{volume}{93}}, \bibinfo{pages}{176601}
  (\bibinfo{year}{2004}{\natexlab{a}}).

\bibitem[{\citenamefont{Silov et~al.}(2004)\citenamefont{Silov, Blajnov,
  Wolter, Hey, Ploog, and Averkiev}}]{silov04}
\bibinfo{author}{\bibfnamefont{A.~Y.} \bibnamefont{Silov}},
  \bibinfo{author}{\bibfnamefont{P.~A.} \bibnamefont{Blajnov}},
  \bibinfo{author}{\bibfnamefont{J.~H.} \bibnamefont{Wolter}},
  \bibinfo{author}{\bibfnamefont{R.}~\bibnamefont{Hey}},
  \bibinfo{author}{\bibfnamefont{K.~H.} \bibnamefont{Ploog}}, \bibnamefont{and}
  \bibinfo{author}{\bibfnamefont{N.~S.} \bibnamefont{Averkiev}},
  \bibinfo{journal}{Appl. Phys. Lett.} \textbf{\bibinfo{volume}{85}}
  (\bibinfo{year}{2004}).

\bibitem[{\citenamefont{Bychkov and Rashba}(1984{\natexlab{a}})}]{bychkov84a}
\bibinfo{author}{\bibfnamefont{Y.~A.} \bibnamefont{Bychkov}} \bibnamefont{and}
  \bibinfo{author}{\bibfnamefont{E.}~\bibnamefont{Rashba}},
  \bibinfo{journal}{JETP Lett.} \textbf{\bibinfo{volume}{39}},
  \bibinfo{pages}{78} (\bibinfo{year}{1984}{\natexlab{a}}).

\bibitem[{\citenamefont{Bychkov and Rashba}(1984{\natexlab{b}})}]{bychkov84b}
\bibinfo{author}{\bibfnamefont{Y.~A.} \bibnamefont{Bychkov}} \bibnamefont{and}
  \bibinfo{author}{\bibfnamefont{E.}~\bibnamefont{Rashba}},
  \bibinfo{journal}{J. Phys. C} \textbf{\bibinfo{volume}{17}},
  \bibinfo{pages}{6039} (\bibinfo{year}{1984}{\natexlab{b}}).

\bibitem[{\citenamefont{Kato et~al.}(2004{\natexlab{b}})\citenamefont{Kato,
  Myers, Gossard, and Awschalom}}]{kato04b}
\bibinfo{author}{\bibfnamefont{Y.~K.} \bibnamefont{Kato}},
  \bibinfo{author}{\bibfnamefont{R.~C.} \bibnamefont{Myers}},
  \bibinfo{author}{\bibfnamefont{A.~C.} \bibnamefont{Gossard}},
  \bibnamefont{and} \bibinfo{author}{\bibfnamefont{D.~D.}
  \bibnamefont{Awschalom}}, \bibinfo{journal}{Science}
  \textbf{\bibinfo{volume}{306}}, \bibinfo{pages}{1910}
  (\bibinfo{year}{2004}{\natexlab{b}}).

\bibitem[{com({\natexlab{b}})}]{commQwire1}
\bibinfo{note}{It must be noted for completeness that within the more rigorous
  microscopic approach in which the effect of impurities is accounted for (see
  for instance Ref.~\onlinecite{aronov89} and \onlinecite{aronov91}) the
  coefficient of the linear contribution to $\mathcal{M}_y(y)$ depends on the
  specific electron impurity potential assumed.}

\bibitem[{\citenamefont{Engel et~al.}(2006)\citenamefont{Engel, Rashba, and
  Halperin}}]{engel06}
\bibinfo{author}{\bibfnamefont{H.-A.} \bibnamefont{Engel}},
  \bibinfo{author}{\bibfnamefont{E.~I.} \bibnamefont{Rashba}},
  \bibnamefont{and} \bibinfo{author}{\bibfnamefont{B.~I.}
  \bibnamefont{Halperin}}, \bibinfo{journal}{cond-mat/0603306}
  (\bibinfo{year}{2006}).

\bibitem[{\citenamefont{Wunderlich et~al.}(2005)\citenamefont{Wunderlich,
  Kaestner, Sinova, and Jungwirth}}]{wunderlich05}
\bibinfo{author}{\bibfnamefont{J.}~\bibnamefont{Wunderlich}},
  \bibinfo{author}{\bibfnamefont{B.}~\bibnamefont{Kaestner}},
  \bibinfo{author}{\bibfnamefont{J.}~\bibnamefont{Sinova}}, \bibnamefont{and}
  \bibinfo{author}{\bibfnamefont{T.}~\bibnamefont{Jungwirth}},
  \bibinfo{journal}{Phys. Rev. Lett.} \textbf{\bibinfo{volume}{94}},
  \bibinfo{pages}{047204} (\bibinfo{year}{2005}).

\bibitem[{\citenamefont{Sih et~al.}(2005)\citenamefont{Sih, Myers, Kato, Lau,
  Gossard, and Awschalom}}]{sih05}
\bibinfo{author}{\bibfnamefont{V.}~\bibnamefont{Sih}},
  \bibinfo{author}{\bibfnamefont{R.~C.} \bibnamefont{Myers}},
  \bibinfo{author}{\bibfnamefont{Y.~K.} \bibnamefont{Kato}},
  \bibinfo{author}{\bibfnamefont{W.~H.} \bibnamefont{Lau}},
  \bibinfo{author}{\bibfnamefont{A.~C.} \bibnamefont{Gossard}},
  \bibnamefont{and} \bibinfo{author}{\bibfnamefont{D.~D.}
  \bibnamefont{Awschalom}}, \bibinfo{journal}{Nature Phys.}
  \textbf{\bibinfo{volume}{1}}, \bibinfo{pages}{31} (\bibinfo{year}{2005}).

\bibitem[{\citenamefont{Sih et~al.}(2006)\citenamefont{Sih, Lau, Myers,
  Horowitz, Gossard, and Awschalom}}]{sih06}
\bibinfo{author}{\bibfnamefont{V.}~\bibnamefont{Sih}},
  \bibinfo{author}{\bibfnamefont{W.~H.} \bibnamefont{Lau}},
  \bibinfo{author}{\bibfnamefont{R.~C.} \bibnamefont{Myers}},
  \bibinfo{author}{\bibfnamefont{V.~R.} \bibnamefont{Horowitz}},
  \bibinfo{author}{\bibfnamefont{A.~C.} \bibnamefont{Gossard}},
  \bibnamefont{and} \bibinfo{author}{\bibfnamefont{D.~D.}
  \bibnamefont{Awschalom}}, \bibinfo{journal}{Phys. Rev. Lett.}
  \textbf{\bibinfo{volume}{97}}, \bibinfo{pages}{096605}
  (\bibinfo{year}{2006}).

\end{thebibliography}
\end{document}